\documentclass{revtex4-1}
\pdfoutput=1
\usepackage{amsmath,amssymb}
\usepackage{graphicx}
\usepackage{subfigure}

\begin{document}
\title{Multiscale Simulation of Entangled Polymer Melt with Elastic Deformation}
\author{Takahiro Murashima}
\email{murasima@cmpt.phys.tohoku.ac.jp}
\affiliation{Department of Physics, Tohoku University, Japan}
\altaffiliation{6-3, Aramaki-Aoba, Aoba-Ward, Sendai, 980-8578, Japan}

\begin{abstract}
To predict flow behavior of entangled polymer melt,
we have developed multiscale simulation composed of Lagrangian fluid particle simulation 
and coarse-grained polymer dynamics simulation.
We have introduced a particle deformation in the Lagrangian fluid particle simulation 
to describe elongation flow at a local point.
The particle deformation is obtained to be consistent with the local polymer deformation.
\end{abstract}


\maketitle

\section{Introduction}
Prediction of entangled polymer melt flow is difficult 
because microscopic polymer dynamics influences the macroscopic flow behavior.
To avoid the difficulties, 
macroscopic fluid dynamics and microscopic polymer dynamics 
are separately considered in conventional ways 
as 'fluid mechanics'~\cite{Bird1} and 'kinetic theory'~\cite{Bird2}.
Macroscopic fluid dynamics of entangled polymer melt 
is numerically solved with computational fluid dynamics (CFD) 
with constitutive equation (CE).
CE is a time-dependent equation of stress representing elasticity and viscosity
coming from microscopic polymer dynamics, 
but many CEs are derived phenomenologically without conformation of polymer chain.
Then, it is difficult to apply to an arbitrary polymer melt 
in the framework of CFD with CE.
Microscopic entangled polymer dynamics 
is solved with molecular dynamics simulation (MD) ~\cite{KremerGrest}
or coarse-grained polymer dynamics simulation 
(CGPD) with polymer chain conformations~\cite{HuaSchieber,Masubuchi,DoiTakimoto,Likhtman,KhaliullinSchieber,
Uneyama,UneyamaMasubuchi}. 
MD or CGPD has microscopic details of polymer chains 
and is applicable to an arbitrary polymer melt. 
To describe flow dynamics of polymer melt (larger than cubic millimeter)
in MD or CGPD (cubic nanometer), 
the number of degrees of freedom in the system becomes more than 10$^{18}$ 
times larger than the original system.
Such a quite large scale simulation is impossible 
even in the world's highest level supercomputer which is accessible 
up to 10$^6$ times scale.
Both macroscopic and microscopic approaches 
still have difficulties in dealing with entangled polymer melt flow.

Hierarchical approaches have been developed 
to compensate for the deficiencies among macroscopic fluid dynamics simulation 
and microscopic molecular dynamics simulation~\cite{LasoOttinger,HuaSchieber1996,Ren,De,
YY2009,YY2010,MT2010,MT2011,MT2012,MYTY2013}.
CONNFESSIT~\cite{LasoOttinger} has pioneered to bridge macroscopic fluid dynamics
and microscopic polymeric system.
This idea is extended 
to liquid crystals employing a different microscopic model~\cite{HuaSchieber1996}.
These pioneering works could not include details of polymer chain conformation.
Rapid progress in computer technology 
has produced heterogeneous multiscale methods (HMM) 
which consist of CFD and MD~\cite{Ren,De,YY2009,YY2010} and have succeeded 
to treat polymer chain conformation in a hierarchical approach.
Then, a Lagrangian multiscale method (LMM) has been developed
in order to manage advection of polymer chain conformation 
which is important for hysteresis in a general polymer melt flow~\cite{MT2010,MT2011,MT2012,MYTY2013}.

LMM consists of Lagrangian fluid particle simulation and CGPD~\cite{MT2010,MT2011,MT2012,MYTY2013}.
A position of a Lagrangian fluid particle represents center of mass of 
a system represented by CGPD.
In a general flow field, polymer chains in a fluid particle is deformed and oriented, 
and then the collective deformation and orientation are observed macroscopically~\cite{MT2012}.

LMM is based on the smoothed particle hydrodynamics method (SPH)~\cite{SPH,MSPH}
and each fluid particle has isotropic density distribution.
When polymer chains in a fluid particle are deformed, 
the density distribution of polymer chains is not isotropic.
This anisotropy is reflected in the normal stress and will cause 
inflow and outflow in Eulerian CFD which uses a fixed mesh.
However, the Lagrangian fluid particle can not express the flow caused 
by the normal stress because of the isotropic density distribution of fluid particle.

To reflect the anisotropy at microscopic level 
to macroscopic density distribution of a fluid particle, 
we need to find a statistical expression of collective deformation of polymer chains.
This expression should be consistent 
with the density distribution of a fluid particle.

In the following section, we discuss on 
the collective deformation of polymer chains in a flow field.
Then, we express an anisotropic density distribution of a fluid particle obtained 
from the collective deformation of polymer chains.
Finally, we summarize this research.

\section{Collective deformation of polymer chains}

In an entangled polymer melt, polymer chains are entangled each other and 
make a complex network structure.
Polymer chains in equilibrium are randomly oriented and isotropically expanded 
with a gyration radius $R_{\rm g}$:
\begin{subequations} 
\label{eq.semiaxis}
\begin{align}
R_{\rm g}^2&=\langle R^2 \rangle = \langle R_1^2 \rangle + \langle R_2^2 \rangle
+ \langle R_3^2 \rangle
=\overline{R}_1^2+\overline{R}_2^2+\overline{R}_3^2, \\
\boldsymbol{R}_k & = \sum_{i=1}^{n_k} \boldsymbol{r}_{i k}, \\
R_{k \alpha}^2 &= \sum_{i=1}^{n_k} r_{i k \alpha}^2 
+ 2 \sum_{i=1}^{n_k} \sum_{j>i}^{n_k} r_{i k \alpha}r_{j k \alpha}, 
\end{align} 
\end{subequations}
where $\boldsymbol{r}_{i k}=r_{i k 1}\boldsymbol{e}_1 +r_{i k 2}\boldsymbol{e}_2
+r_{i k 3}\boldsymbol{e}_3$ is the $i$-th segment vector
in the $k$-th polymer chain in the system with $N$ polymer chains,
$\boldsymbol{e}_{\alpha} (\alpha=\{1, 2, 3\})$ are orthonormal basis,
and the brackets $\langle \cdot \rangle$ express the average over $N$ polymer chains.
$\overline{R}_{\alpha}=\sqrt{\langle R_{\alpha}^2 \rangle}$ represent semiaxes of
an ellipsoid projected onto the orthonormal basis vector $\boldsymbol{e}_{\alpha}$.
At equilibrium, $\overline{R}_1=\overline{R}_2=\overline{R}_3 \equiv R_{\rm g}$, and then we can describe an isotropic sphere.

In a flow field, polymer chains in a complex network structure are deformed 
and oriented according to the flow history~\cite{MT2011}. 
The distribution of polymer chains are not isotropic unlike an equilibrium
state.
However, this anisotropy can be characterized by the orthonormal basis
$\boldsymbol{e}_{\alpha}$.

The stress tensor $\boldsymbol{\sigma}$ is originated in the
anisotropy of the polymer chain conformations:
\begin{subequations} 
\begin{align}
\sigma_{\alpha \beta} & = \frac{\sum_{k=1}^{N} \sigma_{\alpha
 \beta}^{k}}{N} = \langle \sigma_{\alpha \beta}^{k} \rangle,\\
\sigma_{\alpha \beta}^{k} &= \sum_{i=1}^{n_k} r_{ik\alpha}
 F_{ik\beta} \propto \sum_{i=1}^{n_k} r_{ik\alpha} r_{ik\beta},
\end{align} 
\end{subequations}
where $F_{ik\beta}=(3k_{\rm B}T/a^2)r_{ik\beta} (\beta=\{1,2,3\})$ represents a tension on
the $i$-th segment vector in the $k$-th polymer chain $r_{ik \beta}$.
The eigen value $\lambda_{\alpha}$ ($\lambda_1 < \lambda_2 < \lambda_3$) and eigen vector $\boldsymbol{v}_{\alpha}$of the stress tensor $\boldsymbol{\sigma}$ 
represent the orientation degree of the polymer chains and the
characteristic direction of the orientation:
\begin{align}
\boldsymbol{\sigma} \boldsymbol{v}_{\alpha} = \lambda_{\alpha}
 \boldsymbol{v}_{\alpha}, \quad \alpha=\{1, 2, 3\}.
\label{eq.eigen}
\end{align}
Note that the eigen value $\lambda_{\alpha}$ 
does not represent the semiaxis $\overline{R}_{\alpha}$.

\begin{figure}
\begin{center}
\subfigure[]{
\includegraphics[width=5cm]{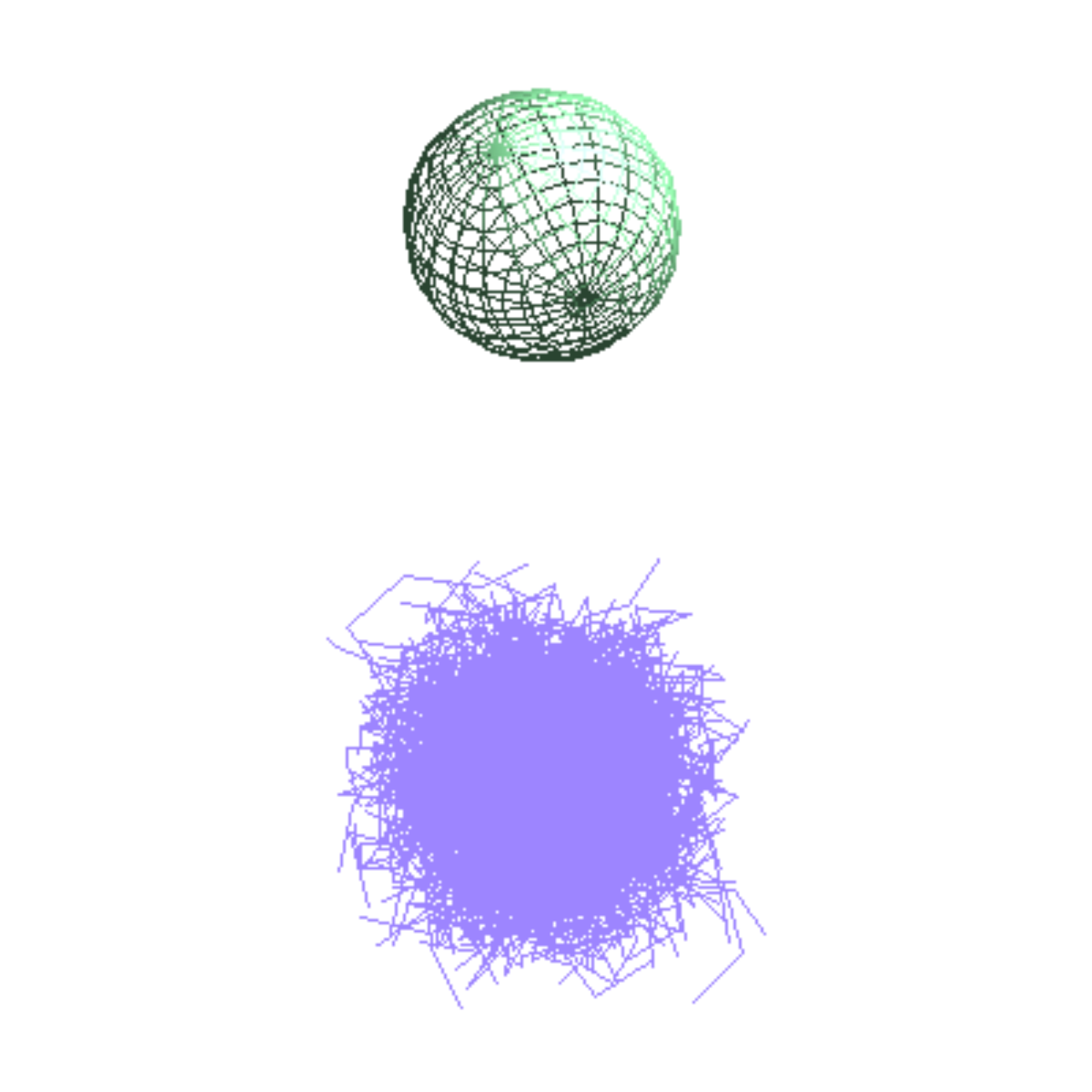}
}
\subfigure[]{
\includegraphics[width=5cm]{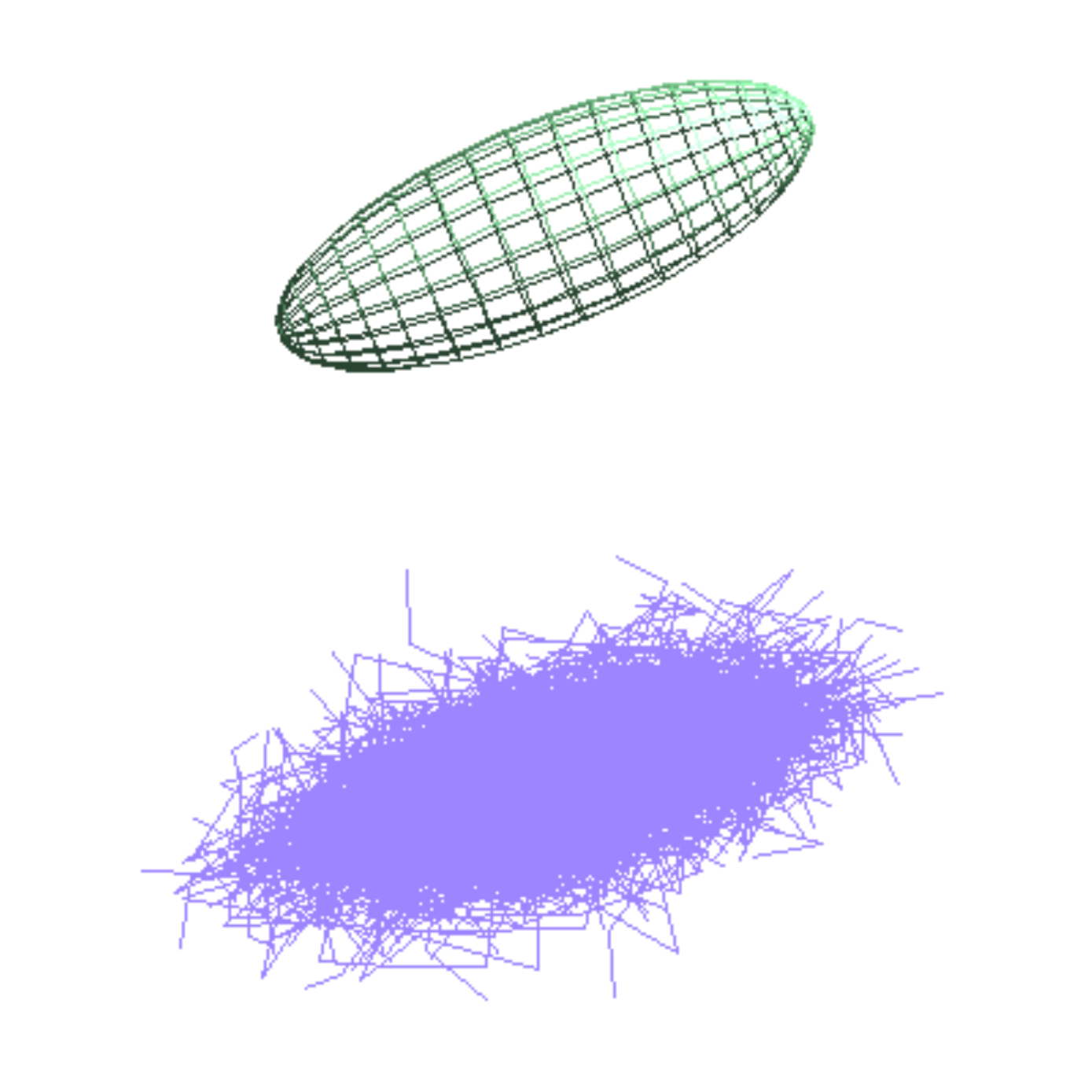}
}
\end{center}
\caption{Equilibrium state (a) and non-equilibrium state (b). 
Top column shows statistical expression of 
collective deformation of polymer chains.
Bottom column shows that one thousand polymer chains are superposed
with fixing the center of mass.}
\label{fig.1}
\end{figure}

Then, we confirm whether the semiaxes $\overline{R}_{\alpha}$ reflect
the anisotropy of the polymer chain distribution or not,
when we choose the eigen vector of the stress tensor
$\boldsymbol{v}_{\alpha}$ as the orthonormal
basis $\boldsymbol{e}_{\alpha}$.
We employ one of the CGPDs, PASTA~\cite{DoiTakimoto,PASTA-CODE}, to produce
an equilibrium state and a non-equilibrium state of entangled polymer
chains.
Simulation condition is as follows: number of polymer chains in a system
$N=1000$, average number of entanglements $Z=\langle n_k \rangle = 10$,
time step $\Delta \tau=0.01$. Unit of time $\tau$ is the relaxation time of entanglement strand. 
Unit of length $a$ is the size of entanglement mesh at equilibrium.
An equilibrium state is obtained after 1,000 $[\tau]$ at rest.
A non-equilibrium state is also obtained after 1,000 $[\tau]$ under
constant shear flow $\dot{\gamma}=0.01 [1/\tau]$.
Figure \ref{fig.1} shows the equilibrium state (a) and the
non-equilibrium state (b).
The top column in Fig. \ref{fig.1} shows the ellipsoids obtained from
Eqs. (\ref{eq.semiaxis}) and (\ref{eq.eigen}), and the bottom
one shows polymer chains superposed with fixing the center of mass.
The sphere in Fig. \ref{fig.1} (a) are described with 
the following parameters obtained from polymer chain distribution:
$\overline{R}_1=1.694360,
\overline{R}_2=1.766702,
\overline{R}_3=1.693777,
\boldsymbol{e}_1=(0.556234, 0.789706, -0.258783),
\boldsymbol{e}_2=(0.763823, -0.363156, 0.533565),
\boldsymbol{e}_3=(-0.327381, 0.494451, 0.805195)$.
Because of the statistical error, the sphere is not a perfect sphere.
The ellipsoid in Fig. \ref{fig.1} (b) are described with the following parameters:
$\overline{R}_1=1.279444,
\overline{R}_2=1.622620,
\overline{R}_3=3.610542,
\boldsymbol{e}_1=(0.380737, -0.921523, 0.076386),
\boldsymbol{e}_2=(-0.034646, 0.068333, 0.997061),
\boldsymbol{e}_3=(0.924034, 0.382264, 0.005911)$.
As shown in Fig. \ref{fig.1}, the obtained ellipsoid at equilibrium 
state is a nearly isotropic sphere, and that at non-equilibrium is
anisotropic.
The both ellipsoids correspond 
to the distribution of polymer chains.

In this section, we have discussed that the collective deformation of polymer chains is
expressed with the anisotropic ellipsoids.
In the next section, we consider how to connect the ellipsoid to the
density distribution of the fluid particle.

\section{Fluid particle deformation}

The anisotropic density distribution of polymer chains, discussed in the
previous section, can be reflected to the macroscopic density
distribution of fluid particle. 
In the conventional SPH~\cite{MSPH}, the density distribution of the $i$-th fluid
particle $\rho_i(\boldsymbol{r})$ is
isotropic and is obtained as follows:
\begin{align}
 \rho_i(\boldsymbol{r})&= m_i W(|\boldsymbol{r}-\boldsymbol{r}_i|,h), \label{eq.density}\\
W(|\boldsymbol{r}|,h)&=
\begin{cases}
\frac{A_d}{(h\sqrt{\pi})^{d}} \left(e^{-|\boldsymbol{r}|^2/h^2} - e^{-4} \right), 
& \quad  |\boldsymbol{r}| \le 2h,\\
 0, & \quad  |\boldsymbol{r}| > 2h,\\
\end{cases}
\label{eq.kernel}
\end{align}
where $m_i$ is the mass of $i$-th fluid particle,
$W$ is the kernel function, $h$ is the width of the kernel $W$.
$A_d$ is the normalization factor in $d$-dimensional space to satisfy
$\int_{|\boldsymbol{r}|\le 2h} {\rm d}\boldsymbol{r}
W(|\boldsymbol{r}|,h) = 1$~\cite{MSPH}.
In the conventional SPH, the fluid is assumed to be a Newtonian fluid
which is an isotropic fluid. However, the polymer melt is not isotropic
as shown in the previous section.
The fluid particle of polymer melt should be anisotropic according to the
distribution of polymer chains. 
The anisotropic density distribution can be produced by the following
kernel:
\begin{align}
W(\{ r_{\alpha} \},\{ \epsilon_{\alpha}\},h)&=
\begin{cases}
\frac{A_d}{(h\sqrt{\pi})^{d}(\Pi_{\alpha=1}^d\epsilon_{\alpha})} \left( \Pi_{\alpha=1}^d
 e^{-r_{\alpha}^2/(\epsilon_{\alpha}h)^2} - e^{-4} \right),
 & \quad
 \sum_{\alpha=1}^{d} \left( \frac{r_{\alpha}}{\epsilon_{\alpha}} \right)^2 \le 4h^2,\\
 0, 
&\quad
  \sum_{\alpha=1}^{d} \left( \frac{r_{\alpha}}{\epsilon_{\alpha}} \right)^2  > 4h^2,\\
\end{cases}
\label{eq.ekernel}
\end{align}
where $\{\epsilon_{\alpha}\}$ is the ratio of semiaxis to the radius of
the sphere and is a dimensionless variable.
Equation (\ref{eq.ekernel}) corresponds to Eq. (\ref{eq.kernel})
when $\epsilon_{\alpha}=1.$
If we assume an affinity to the density distribution of a fluid particle
and the density distribution of the polymer chains in the fluid
particle,
$\epsilon_{\alpha}\equiv \overline{R}_{\alpha}/R_{\rm g}$.
This assumption is reasonable because the macroscopic resolution is
nearly equal to $h$ in this multiscale simulation and $h$ is much larger
than the polymer chain length in the fluid particle.
Higher order deformation at microscopic level is ignorable in the
multiscale simulation.

\begin{figure}
\begin{center}
\subfigure[]{
\includegraphics[width=5cm]{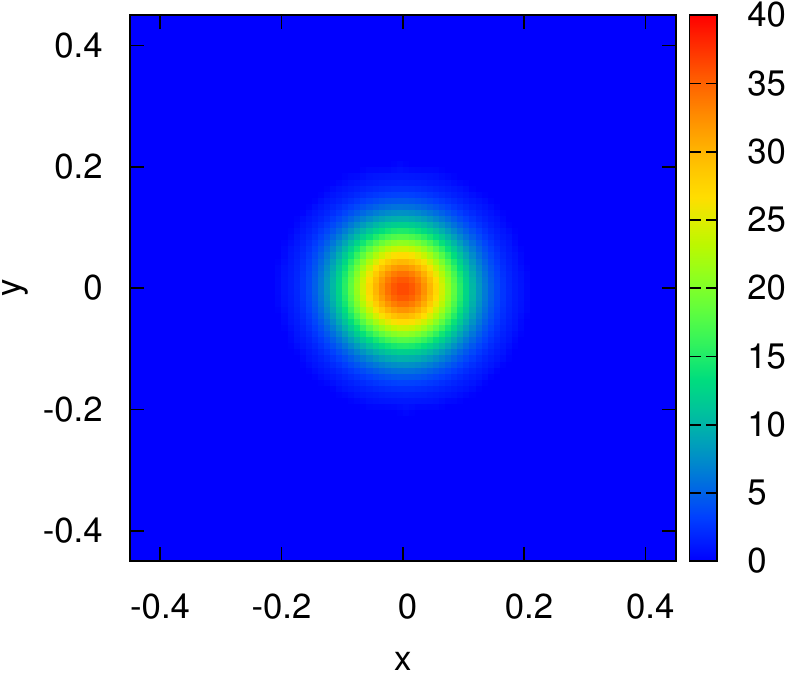}
}
\subfigure[]{
\includegraphics[width=5cm]{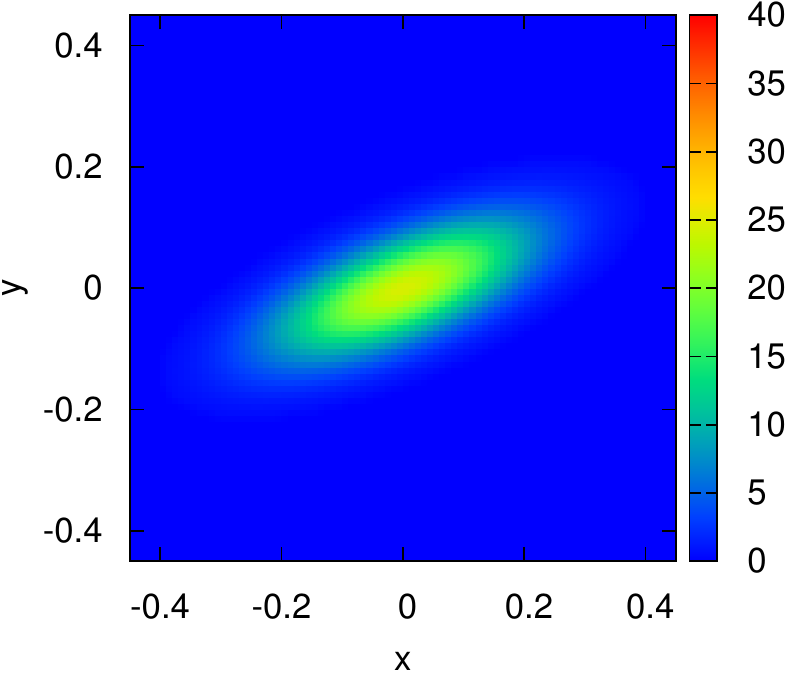}
}
\end{center}
\caption{Density distribution of a fluid particle at equilibrium state
 (a) and non-equilibrium state (b) on $z=0$ plane. }
\label{fig.2}
\end{figure}

Figure \ref{fig.2} shows the density distribution of a fluid particle on $z=0$ plane,
obtained from Eqs. (\ref{eq.density}) and (\ref{eq.ekernel}).
The mass center of fluid particle is fixed to be $(0, 0, 0)$.
Figures \ref{fig.2} (a) and (b) correspond to Figs \ref{fig.1} (a) and (b),
respectively. The following parameters are assumed: 
$m=1.0[{\rm M}]$, $h=0.1[{\rm L}]$, $d=3$, $A_3=1.18516$~\cite{MSPH}, where ${\rm M}$ is 
macroscopic mass unit and ${\rm L} (\ne a)$ is macroscopic length unit.

\section{Summary}

We have considered the collective deformation of polymer chains in a flow field 
and derived a statistical way to describe the collective deformation by an ellipsoid.
Then, we have connected this microscopic anisotropy to the macroscopic fluid particle.
The anisotropy of fluid particle produces heterogeneity in the density distribution 
and will causes a flow to recover a uniform density distribution.
This flow can produce 'Barus effect' or 'die swell'~\cite{book.rheo} 
which was not managed in the previous works~\cite{MT2010,MT2011,MT2012,MYTY2013}.
We expect that LMM will be improved with the techniques shown above
and can solve an arbitrary polymer melt flow.

\section*{Acknowledgments}
The computation in this work has been done using 
the facilities of the Supercomputer Center, 
the Institute for Solid State Physics, the University of Tokyo, 
and the supercomputer of ACCMS, Kyoto University.
This work was supported by JSPS KAKENHI (Grant Number 23340120 and 24350114),
JSPS Core-to-Core program 'Non-equilibrium dynamics of soft matter and information',
and the National Institute of Natural Science (NINS)
Program for Cross-Disciplinary Study.

\end{document}